%% file: paper.tex
\documentclass{article}
\usepackage{spconf,amsmath,graphicx}
\usepackage{multirow}
\usepackage{xcolor}
\usepackage{amsfonts}
\usepackage{subfig}
\usepackage{amssymb}
\usepackage{tabularx}
\usepackage{multirow}
\usepackage{cite}

\usepackage{bbm}
\usepackage{etoolbox}
\usepackage{siunitx}
\usepackage{booktabs}

\title{Towards Better Meta-Initialization with Task Augmentation for Kindergarten-aged Speech Recognition}
\name{Yunzheng Zhu, Ruchao Fan, and Abeer Alwan}
\address{Department of Electrical and Computer Engineering, University of California, Los Angeles, USA}
\begin{document}
%
\maketitle
%
\begin{abstract}
Children's automatic speech recognition (ASR) is always difficult due to, in part, the data scarcity problem, especially for  kindergarten-aged kids. When data are scarce, the model might overfit to the training data, and hence good starting points for training are essential. Recently, meta-learning was proposed to learn model initialization (MI) for ASR tasks of different languages. This method leads to good performance when the model is adapted to an unseen language. However, MI is vulnerable to overfitting on training tasks (learner overfitting). It is also unknown whether MI generalizes to other low-resource tasks. In this paper, we validate the effectiveness of MI in children's ASR and attempt to alleviate the problem of learner overfitting. To achieve model-agnostic meta-learning (MAML), we regard children's speech at each age as a different task. In terms of learner overfitting, we propose a task-level augmentation method by simulating new ages using frequency warping techniques. Detailed experiments are conducted to show the impact of task augmentation on each age for kindergarten-aged speech. As a result, our approach achieves a relative word error rate (WER) improvement of 51\% over the baseline system with no augmentation or initialization.

\end{abstract}
\section{Introduction}
\label{sec:intro}
Child ASR is a challenging problem, in part, because of the lack of large child speech databases. This is especially true for kindergarten-aged children \cite{yeung2021fundamental}, even though ASR technology for such young kids might be helpful in literacy instruction and assessment. The main problem with insufficient training data is that the resulting acoustic model does not generalize well because of optimizing to local minima during training. A common approach used to address this problem is data augmentation using techniques such as SpecAug \cite{park2019specaugment}, speed perturbation \cite{ko2015audio}, and VTLP \cite{jaitly2013vocal}. 



Another possible solution for this problem is model-agnostic meta-learning (MAML) \cite{finn2017model,nichol2018first}. Meta-learning allows for fast adaption from different tasks to an unseen task, and is referred to as meta-initialization (MI) \cite{DBLP:conf/iclr/AntoniouES19,wang2021improved}. The idea is to learn a good model initialization from different training tasks. It has been shown to be  effective in cross-accent \cite{winata20_interspeech} and multi-lingual ASR \cite{hsu2020meta} as well as in other fields such as computer vision \cite{NIPS2017_cb8da676}, neural machine translation \cite{gu-etal-2018-meta}, and speaker adaptive training \cite{klejch2018learning}. However, MI is also vulnerable to learner overfitting \cite{yao2021improving,ni2021data}, which
happens when the model overfits to the training tasks and is unable to generalize to the testing task.

To address the issue of learner overfitting, several task augmentation based mechanisms were proposed. Liu et al. treated each rotation of an image as a new task for image classification tasks \cite{liu2020task}, and Murty et al. proposed \textit{DRECA} that uses latent reasoning categories to form new tasks for natural language processing tasks \cite{murty2021dreca}. To our knowledge, no study has addressed the issue of learner overfitting in ASR before.

In this paper, we discover how meta-learning and task-based augmentation algorithms can apply to kindergarten children's ASR. In MI, the tasks are defined according to the development of children's vocal tract because it varies by the child's age. 
Although a promising improvement is observed with the MI for kindergarten-aged speech, learner overfitting occurs. To alleviate learner overfitting, we propose a task augmentation mechanism for children's ASR by simulating new tasks using speed perturbation, and spectral shifting-based data augmentation methods, VTLP, because of the characteristics of each task (vocal tract differences). 

The remainder of this paper is organized as follows: Section 2 presents the meta-initialization and task augmentation approaches for the low resource kindergarten-aged ASR. Section 3 describes the experimental setup, followed by results and discussion in Section 4. Section 5 concludes this paper.



\section{Method}
\label{sec:method} 
For a data sufficient task, traditional machine learning can generalize well for in-domain data using random parameter initialization. However, when data are scarce, random initialization might overfit to the training data easily, and hence good starting points for training are essential for better model generalization. Previously, it has been shown that supervised pre-training can provide a good starting point for training in low resource tasks \cite{tong2017transfer}. As mentioned earlier, the aim of meta-learning application is to provide good initialization for low-resource tasks by quickly adapting the knowledge learned from the different available tasks to the unseen task, and is referred to as meta-initialization (MI). However, meta-initialization can be at risk of overfitting to the training tasks; this is referred to as learner overfitting \cite{NEURIPS2020_3e5190ee}. In this section, we show how to use MI for ASR of children's speech and describe the proposed task-level augmentation method for solving the learner overfitting problem.

\subsection{Meta-initialization (MI)}

\label{ssec:metainit}
Meta-learning is defined as a "learning to learn" method where the goal is to design a strategy to better choose a system's hyperparameters and learning algorithm. Learning model initialization, or meta-initialization (MI) is also one of the most important components in meta-learning. Suppose we have a set of training tasks $\mathbb{G} = \{G_1, G_2, \ldots, G_i, \ldots, G_n\}$ and a target test task $T$. The idea is to simulate the adaptation stage during training and minimize an objective function. Note that the objective function is based on the adapted model so that the model before adaptation can be regarded as a good model initialization for the adaptation stage. For each training task $G_i$, the data are split into a support set $G_i^{sup}$ that is used in the inner loop for the adaptation stage, and a query set $G_i^{que}$ for evaluating the effectiveness of the model after the task's adaptation stage. A better initial model before adaptation leads to better performance. The loss function based on the query set is used in the outer loop to calculate the final objective function.

Suppose that the model parameters in the inner-loop are $\theta_j$ at step $j$, the audio samples in the support set of each training task $G_i^{sup}$ are used to simulate the adaptation stage. The model is updated as follows:
\begin{flalign}
    &\phi_{ji} = \theta_j - \alpha\triangledown_{\theta_j}\textit{L}(f(X_i^{sup};\theta_j), Y_i^{sup})
    \label{eq:1}
\end{flalign}
where $X_i^{sup}$ and $Y_i^{sup}$ are data samples and corresponding labels in the support set of task $i$, respectively. $\phi_{ji}$ is the model parameter updated for task $i$ and step $j$. $f$ is the forward computation of the model. $L$ is the cross-entropy loss used in acoustic modelling, and $\alpha$ is the learning rate for the inner-loop optimizer. $\triangledown$ is the nabla operator for computing the gradient of $\theta_{j}$.

In the outer-loop, we quantify how the adaptation behaves in the inner loop by a summation over the loss function for the query set of each task. The summation is referred to as the meta-objective function:
\begin{flalign}
    \sum_{{G_{i}}} \textit{L}(f(X_i^{que};\phi_{ji}), Y_i^{que})
\end{flalign}
where $X_i^{que}$ and $Y_i^{que}$ are data samples and corresponding labels in the query set of task $i$, respectively. By minimizing the above objective function with respect to $\theta_j$, we can find a model that is suitable for adaptation, and hence the model can be regarded as a good initialization. After the optimization, which is based on the inner loop, is completed (Eq.\ref{eq:1}), the initialization would be the focus of the algorithm.
\begin{flalign}
    \theta_{j+1} \leftarrow \theta_j - \beta\triangledown_{\theta_j}' \sum_{{G_{i}}}\textit{L}(f(X_i^{que};\phi_{ji}), Y_i^{que}))
\end{flalign}
where $\beta$ is the learning rate for the outer-loop optimizer, and $\triangledown_{\theta_j}'$ indicates that only first-order MAML \cite{nichol2018first} is used since the second-order derivative is computationally expensive and it does not affect the results significantly. 
After enough training steps, $N$, the final model $\theta_N$ is regarded as the learned initialization for the unseen test task.

\subsection{Age-based Task Augmentation for MI} 
\label{ssec:taskda}
Different from overfitting in traditional machine learning algorithms, there are two other overfitting problems in MI, which are memorization overfitting \cite{Yin2020Meta-Learning} and learner overfitting. The memorization overfitting happens when the $\theta_{j+1}$ memorizes all tasks and does not rely on support sets for inner-loop adaptation. The learner overfitting happens when the $\theta_{j+1}$ is unable to generalize well on the test task $T$. The memorization can be well mitigated by randomly sampling the support set and query set at each step during training since each sample has the opportunity to participate in either inner loop updates or outer loop updates. In terms of the learner overfitting, a common strategy is to use task augmentation to increase the model generalization for the test task. However, task augmentation has not been explored in ASR, to our knowledge, before.

We propose an age-based task augmentation framework to alleviate the problem of learner overfitting in kindergarten-aged speech recognition. The higher degree of inter-speaker variability of children speech is mainly due to different growth patterns of children. These differences result in shifts in the fundamental frequency ($F_0$) and formant frequencies ($F_1, F_2, F_3$, etc.) in kids' speech as they grow. Hence, we perform the augmentation by simulating new tasks of children's speech using time and frequency warping techniques, such as VTLP and speed perturbation. For example, the task for each age $G_i$($G_i^{1.0}$) is augmented with two new tasks with two warping factors 0.9 ($G_i^{0.9}$) and 1.1 ($G_i^{1.1}$). We compare the two techniques in Section \ref{sec:expsetup}.

\section{Experimental Setup}
\label{sec:expsetup}
Experiments are conducted using the Kaldi toolkit \cite{povey2011kaldi} for feature extraction and WFST-based decoding and Pykaldi2 \cite{lu2019pykaldi2} for acoustic model training.

\subsection{Database}
\label{ssec:database}
The database for the experiments is the scripted part of OGI Kids' Speech Corpus \cite{shobaki2000ogi}. The Corpus contains kids speech in eleven age groups from kindergarten, grade 1 (G1) to grade 10 (G10). Each age group has approximately 100 speakers saying single words, sentences, and digit strings. The dataset is randomly split into 70 $\%$ training data, 8 $\%$ development data, and 22 $\%$ test data without speaker overlap for each age as in \cite{fan2021bi}. The kindergarten-aged task is regarded as the meta-testing task for fine-tuning. G1 speech data, which corresponds to the closest age to kindergarten speech, are used for the validation task in meta-learning. Other tasks with kids speech from G2 to G10 are used as the training tasks, which is similar to pre-training, for obtaining a model initialization. For meta-training and meta-validation tasks, training and development sets are combined for sampling the support and query sets. Note that the training data for kindergarten-aged speech is approximately 4 hours and the training data for the meta-initialization stage is about 45 hours.

\subsection{Acoustic Model Setup}
\label{ssec:amsetup}
First, an HMM-GMM model is trained with all the data in the meta-training tasks to obtain frame-level alignment for the DNN-based acoustic model training. 80-dimensional log-mel-filter bank features are extracted every 10 ms with a 25 ms window. An additional frame of features after each frame is appended to form a 160-dimensional input \cite{sak2015fast}. The model has 4 BLSTM layers with 512 hidden units in each direction. The last layer transforms the outputs of BLSTM to a probability distribution of the 1360 states from the HMM model. For the baseline and adaptation of kindergarten-aged task, the training process takes 15 iterations. An Adam optimizer with a multi-step scheduler is applied, where the learning rate is initially set to $1e^{-5}$ for the first two iterations and decayed with a ratio of $0.1$ till the last iteration.

\subsection{Meta-initialization Setup}
In MI, the support set and query set are randomly sampled with a batch size of 16 for each age of G2 to G10 during training. The same frame-level alignment and BLSTM model configuration are used as mentioned in Section \ref{ssec:amsetup}.

The number of iterations for MI training is empirically set to 6,800. Separate optimizers are applied to the outer-loop and inner-loop optimization. The inner loop uses a SGD optimizer with a fixed learning rate of $2e^{-4}$. The outer loop uses an Adam optimizer with a multi-step scheduler, where the learning rate is stabilized to $2e^{-4}$ for the first 2,000 iterations and decayed with a ratio of $0.15$ to $3e^{-5}$ till the last iteration. All the parameters trained from MI are used as the initialization for the training in the adaptation stage.

\subsection{Augmentation Setup}
\label{ssec:dasetup}
For age-based task augmentation during the MI stage, speed perturbation and vocal tract length perturbation (VTLP) are used with the warping factors of 0.9, 1.0, and 1.1, according to our preliminary results \cite{gretter21_interspeech,wang21m_interspeech}, and hence the number of tasks is increased by 3 folds. Thus, we adopt an online augmentation mechanism where at each iteration the warping factor is randomly selected from (0.9, 1.0, 1.1).

During the adaptation stage, speed perturbation and VTLP are used with same warping factors (0.9, 1.0, 1.1) as task augmentation. 
For SpecAug, a maximum width of 5 frequency channels are masked twice, and a maximum width of 8 time channels are masked twice as well. The width of the frequency and time channel are chosen empirically. 


\section{Results and Discussion}
\label{sec:resdis}
An HMM-DNN hybrid system with BLSTM modelling is used as our baseline. As shown in Table \ref{tab:tab1}, the development and test set of kindergarten speech have a WER of 53.17\% and 55.01\%, respectively, without any prior knowledge. The baseline WER is similar to that reported in \cite{gretter21_interspeech} for a small size (5 hours) kids dataset.

\input{Tables/Table1}

\subsection{MI and Task Augmentation}
\label{ssec:MITA}
The results of MI and the proposed task augmentation methods are shown in Table \ref{tab:tab1}. As we can observe from the table, using data augmentation (Data Aug) strategies can improve the performance over baseline. The relative improvement in WER for speed perturbation (SP) and VTLP is around 20\%. When training with an initialization through meta-learning, the WER of the kindergarten-aged test set is decreased from 55.01\% to 30.68\%, a larger relative WER improvement than the data augmentation strategies. For a fair comparison, we used the supervised pre-training method (SPT) to directly train the acoustic model with data from G2-G10 as the starting point. We can see from the table that MI is slightly worse than SPT on the test set.  



The proposed task augmentation methods are used to address the overfitting problem and we observe a significant improvement over the MI without augmentation. From Table \ref{tab:tab1}, we found that SP is better than VTLP as a method to simulate new tasks. For a fair comparison, we also experimented with augmentation that is not task dependent. In raw augmentation (Raw Aug), warping is applied to the original data. The results validate the effectiveness of the proposed task augmentation (Task Aug) method, which achieves a WER of 27.5\% on the kindergarten test set. SpecAug is not used in task augmentation since it randomly masks out time or frequency channels. Such masking is not consistent for the data in one task that is regarded as a new task after augmentation.


\input{Figures/Figure1}

\subsection{The Impact of the Augmented Tasks}
\label{ssec:dacomp}

The task augmentation in Table \ref{tab:tab1} is using speech data from all ages in the training set to augment a new ASR task. To obtain an insight into the impact of the augmented tasks on WER performance, we add the number of augmented tasks incrementally according to age. For example, as shown in Fig.\ref{fig:fig1}, the number of tasks is added in either an increasing order (from G2 to G10), or a decreasing order (from G10 to G2). Our goal is to investigate which subset of the data is more important for the augmentation.

Since SP outperforms VTLP in the previous experiments, SP is explored. As shown in Fig.\ref{fig:fig1}, 
including more augmented tasks in either the forward order or reverse order results in improved performance. However, the reverse order generally performs worse than the forward order by 1\% WER for the kindergarten-aged test set, which means creating new tasks that is similar to the target task is effective in addressing the learner overfitting problem. With all tasks being augmented, the final performance has a 10\% relative WER improvement over MI without the task augmentation. 


\input{Tables/Table3}

\subsection{Data Augmentation for Adaptation}
The task we are focusing on is a low-resource one (kindergarten ASR). Hence, data augmentation methods are further used during the adaptation stage of the kindergarten-aged task. SP, SpecAug and VTLP are compared in the experiments. The results are shown in Table \ref{tab:tab3}. Although all three strategies can improve the performance on the development set, only SpecAug achieves a slightly better performance on the test set. The reasons why VTLP and SP did not achieve better results will be explored in the future work. 

\section{Conclusion}
\label{sec:conclusion}
In this paper, to deal with the data scarcity of children's speech, particularly kindergarten-aged, meta-initialization is used to find a good starting point for training the acoustic model. To mitigate the overfitting problem in meta-initialization, particularly learner overfitting, an age-based task augmentation mechanism is proposed to simulate new ages using time and frequency warping techniques. The data augmentation strategies using speed perturbation and VTLP that are also used in the task augmentation stage are not helpful in the adaptation stage. SpecAug used in the adaptation stage resulted in small WER improvement, and the final system achieved a 51\% relative WER improvement over the baseline (no augmentation and no adaptation). In the future, we will explore the use of the proposed algorithm in other low-resource tasks for both adults and children's ASR.

\section{Acknowledgement}
\label{sec:ack}
This work was supported in part by National Science Foundation (NSF).

\vfill\pagebreak



\bibliographystyle{IEEE}
\bibliography{paper}
\end{document}

%% file: Tables/Table1.tex
\begin{table}[]
    \caption{\% Word error rate (WER) for Data Augmentation (Data Aug) mechanisms on baseline system, meta-initialization (MI), and the proposed task augmentation (Task Aug) mechanisms for MI with vocal tract length perturbation (VTLP) and speed perturbation (SP) on the Kindergarten-aged development and test sets. SPT stands for supervised pre-training. Raw Aug stands for augmentation within each task without creating new tasks. }
    \centering
    \begin{tabular}{lcccc}
        \hline
        \multirow{2}{*}{Model} & Data Aug & MI Aug & \multirow{2}{*}{Dev} & \multirow{2}{*}{Test} \\
         & Type & Type & &  \\
        \hline\hline
        Baseline & - & - & 53.17 & 55.01 \\
        \multirow{3}{*}{+ Data Aug} 
        & SP & - & 46.13 & 43.75 \\
        & VTLP & - & 45.42 & 46.05 \\
        & SpecAug & - & 56.69 & 53.70 \\
        + SPT \cite{tong2017transfer} & - & - & 36.27 & 29.06 \\
        + MI & - & - & 35.21 & 30.68 \\
        \hspace{1mm} \multirow{2}{*}{+ Raw Aug} 
         & - & SP & 36.62 & 28.00 \\
         & - & VTLP & 36.27 & 30.06 \\
        \hspace{1mm} \multirow{2}{*}{+ Task Aug} 
        & - & SP & \textbf{34.86} & \textbf{27.50} \\
         & - & VTLP & \textbf{34.86} & 29.06 \\
        \hline
    \end{tabular}
    
    \label{tab:tab1}
\end{table}

%% file: Figures/Figure1.tex
\begin{figure}[tb]

\begin{minipage}[b]{1.0\linewidth}
  \centering
  \centerline{\includegraphics[width=8.5cm]{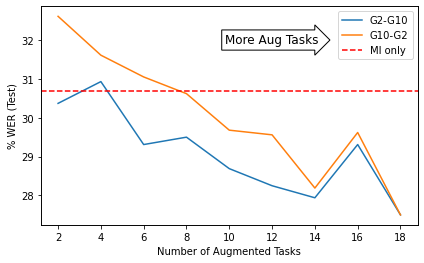}}
\end{minipage}
\caption{\% WER Results of task augmentation mechanism using speed perturbation (SP) versus the number of augmentation tasks for MI on the Kindergarten test set. The tasks are added either from G2 to G10 (in blue), or from G10 to G2 (in orange). The dashed line (in red) is MI without any task augmentation mechanism.}
\label{fig:fig1}
\end{figure}

%% file: Tables/Table3.tex
\begin{table}[]
    \caption{\% Word error rate (WER) for data augmentation during the adaptation stage with SpecAug, vocal tract length perturbation (VTLP), and speed perturbation (SP) on the Kindergarten development and test sets. }
    \centering
    \begin{tabular}{lcc}
        \hline
        Aug Type (in adaptation stage) & Dev & Test \\
        \hline\hline
        No Aug & 34.86 & 27.50 \\
        SpecAug & 32.75 & \textbf{27.01} \\
        VTLP & \textbf{32.39} & 28.13 \\
        SP & 33.45 & 27.75 \\
        \hline
    \end{tabular}
    
    \label{tab:tab3}
\end{table}